\documentclass[useAMS,usenatbib,epsfig, a4paper]{mn2e}
 
\usepackage{float}
\usepackage{epsfig}
\usepackage{subfigure}
\usepackage{epstopdf}
\usepackage{verbatim}
\usepackage{nicefrac}
\usepackage{hyperref}
\usepackage{url}

\title[Power spectra from peculiar velocity catalogues]
    {Power Spectrum Estimation from Peculiar Velocity Catalogues}
\author[E.~Macaulay et al.]
{E.~Macaulay$^1$\thanks{email: \href{mailto:edward.macaulay@astro.ox.ac.uk}{\nolinkurl{edward.macaulay@astro.ox.ac.uk}}}, H.~A.~Feldman$^{2}$, P.~G.~Ferreira$^1$,  A.~H.~Jaffe$^{3}$,\newauthor S.~Agarwal$^{2}$, M.~J.~Hudson$^{4,5}$, R.~Watkins$^{6}$\\ 
$^1$Astrophysics, University of Oxford, Denys Wilkinson Building, Keble 
Road, Oxford OX1 3RH, United Kingdom.\\
$^2$ Department of Physics and Astronomy, University of Kansas, Lawrence, KS66045, USA. \\
$^3$ Department of Physics, Blackett Laboratory, Imperial College, London SW7 2AZ, United Kingdom \\
$^4$ Department of Physics and Astronomy, University of Waterloo, Waterloo, ONN2L3G1, Canada. \\
$^5$ Perimeter Institute for Theoretical Physics, 31 Caroline St. N., Waterloo, ON, N2L2Y5, Canada. \\
$^6$ Department of Physics, Willamette University, Salem, OR97301, USA. \\
}
\date{\today}

\pubyear{2012}

\newcommand{\lcdm}{$\Lambda$CDM }
\newcommand{\kms}{km s$^{-1}$ }
\newcommand{\rUnit}{$h^{-1}$Mpc }

\begin{document}
\maketitle

\begin{abstract}
The peculiar velocities of galaxies are an inherently valuable cosmological probe, providing an unbiased estimate of the distribution of matter on scales much larger than the depth of the survey.  Much research interest has been motivated by the high dipole moment of our local peculiar velocity field, which suggests a large scale excess in the matter power spectrum, and can appear to be in some tension with the \lcdm model.   We use a composite catalogue of 4,537 peculiar velocity measurements with a characteristic depth of 33 \rUnit to estimate the matter power spectrum.  We compare the constraints with this method, directly studying the full peculiar velocity catalogue, to results from \protect\cite{2011MNRAS.tmp..391M}, studying minimum variance moments of the velocity field, as calculated by \protect\cite{2009MNRAS.392..743W} and \protect\cite{2009arXiv0911.5516F}.  We find good agreement with the \lcdm model  on scales of $k>0.01$ $h$ Mpc$^{-1}$.  We find an excess of power on scales of $k<0.01$ $h$ Mpc$^{-1}$, although with a 1$\sigma$ uncertainty which includes the \lcdm model.  We find that the uncertainty in the excess at these scales is larger than an alternative result studying only moments of the velocity field, which is due to the minimum variance weights used to calculate the moments.  At small scales, we are able to clearly discriminate between linear and nonlinear clustering in simulated peculiar velocity catalogues, and find some evidence (although less clear) for linear clustering in the real peculiar velocity data.
  \end{abstract}

\begin{keywords}
cosmology: large scale structure of the universe -- cosmology: observation -- cosmology: theory -- galaxies: kinematics and dynamics -- galaxies: statistics
\end{keywords}

\section{Introduction \& Background}
The peculiar velocities of galaxies are a powerful cosmological probe, which directly traces the underlying dark matter distribution (independent of galaxy bias) and is also sensitive to scales much larger than the size of the survey.  Recent interest in peculiar velocities has been driven by the high dipole moment of our local velocity field, which can appear to be in some tension with the \lcdm model.  The \lcdm model indicates that we should expect a peculiar velocity dipole of magnitude around 100 km s$^{-1}$, although many independent peculiar velocity surveys show evidence for a bulk flow at low redshift of around 400 \kms in the direction $l=280$ $b=10$ degrees \citep{2004MNRAS.352...61H,2009MNRAS.392..743W,2009arXiv0911.5516F,2010arXiv1010.4276M}.  However, \cite{0004-637X-736-2-93} find evidence for a flow more commensurate with \lcdm of around 260 \kms (in a similar direction as the other bulk flow results).  

In more tension with $\Lambda$CDM are significantly higher bulk flows at redshift $\sim$0.25, which curiously appear to be in the same direction as the low redshift bulk flows.  \cite{2011arXiv1106.5791A} found evidence for an extremely high bulk flow of around 4000 km s$^{-1}$ at redshift $\sim$0.3, in a similar direction to other bulk flows.  \cite{2041-8205-712-1-L81} find a bulk flow of around 1000 km s$^{-1}$ extending to $z\simeq0.2$ from kinematic Sunyaev-Zel'dovich measurements.  If substantiated, these bulk flows would indicate a non convergence of the peculiar velocity dipole, and present a serious challenge to the assumptions of isotropy and homogeneity.  However, these bulk flow measurements are far from as robust as the directly measured peculiar velocity measurements we consider here; \cite{2011arXiv1106.5791A} note that their result may be entirely due to systematic effects.  In this work, we will only consider the direct peculiar velocity measurements at low redshift.

In addition to the high dipole moment, there also appears to be a low shear of the velocity field \citep{1995ApJ...455...26J,2009arXiv0911.5516F}, which indicates that the density contrast responsible for the velocity dipole is on extremely large scales.  At the depths of the peculiar velocity surveys (up to 100 $h^{-1}$Mpc), this suggests an excess density contrast on scales $\sim1$ $h^{-1}$Gpc.  The volume of space probed by galaxy redshift surveys (e.g. \citealt{2005MNRAS.362..505C}; \citealt{2010MNRAS.404...60R} and \citealt{2010MNRAS.406..803B}) is currently too small to robustly constrain clustering on these scales, although \cite{2010arXiv1012.2272T} found evidence for excess large scale power in the MegaZ photometric redshift survey.  Measurements of the CMB can probe anisotropies on these large scales \citep{2011arXiv1105.4887H}, although to compare to results at low redshift, the growth of these anisotropies must be assumed, which depends on the cosmological model.

There are many possible explanations for the high dipole moment, ranging from systematic effects to more exotic explanations invoking extended cosmology or modified gravity.  \cite{2004MNRAS.352...61H} considered systematic effects in the SMAC (Streaming Motions of Abell Clusters) peculiar velocity survey, such as a dipole variation in the velocity dispersion, galactic extinction, and calibration across different observations.  They found that systematic effects could account for at most half of the high dipole moment of the SMAC survey, leaving a dipole moment which is still at least three times higher than the \lcdm expectation.  There are many theoretical possibilities to extend the \lcdm model to produce a large scale excess of power and a peculiar velocity dipole, such as modifications to gravity, \citep{2009arXiv0908.2903A,2009PhRvD..80f4023K}, dark energy clustering  \citep{2011JCAP...09..005P}, vorticity, \citep{2010EPJC...69..581P} or `tilted' universes \citep{2009JCAP...02..006M,2041-8205-712-1-L81,2010arXiv1010.4276M}.  In this work we do not address any theoretical explanation in particular, but by using the peculiar velocity data to constrain the power spectrum in model independent band-powers we aim to provide results which will be useful to constrain a range of explanations for the high dipole.

\cite{1995ApJ...455...26J}, \cite{1996ApJ...458..419K}, \cite{1997ApJ...486...21Z}, \cite{1997ApJ...479..592K}, \cite{2001MNRAS.326..375Z} and \cite{2001ApJ557102S}  have used peculiar velocity surveys to infer the matter power spectrum directly at $z=0$.  More recently in \cite{2011MNRAS.tmp..391M}, we inferred the underlying power spectrum from moments of the peculiar velocity field from \cite{2009MNRAS.392..743W} and \cite{2009arXiv0911.5516F}, specifically in the context of understanding the high bulk flow.  One of our main findings was that the excess of power indicated by the anomalously high dipole moment was dramatically reduced when the shear and octupole moments were also included.   This leads us to ask if the correspondence with the \lcdm model would be improved if we were to hypothetically include higher still moments of the velocity field.  That is the motivation for this work, although we take the approach used by \cite{1995ApJ...455...26J} to analyse peculiar velocity catalogues directly, without compressing the data into moments.  

Recently, \cite{2009MNRAS.400.1541A} applied a similar formalism to constrain the modified gravity parameter $\gamma$.  Similarly \cite{2010arXiv1010.4276M} used a similar formalism to fit for parameters of the dipole moment and velocity dispersion parameter.  In both cases, a fiducial \lcdm power spectrum was assumed, and additional parameters of interest were allowed to vary.  The key difference between those papers and this work is that here we treat the underlying power spectrum as a set of free parameters, as opposed to fitting extra parameters for additional effects beyond a fixed power spectrum.  In this way, we obtain new constraints on the power spectrum which are independent of the fiducial cosmology.

\section{Method}
\label{Method}

In this paper we consider two distinct methods to relate peculiar velocity measurements to large scale structure: A maximum likelihood based approach to analyse a full peculiar velocity catalogue (the `catalogue' method), and an alternative approach studying minimum-variance moments of the velocity field (the `moments' method). 

As well as an apparent radial velocity due to the Hubble flow, galaxies also have a peculiar velocity towards local over-densities of matter.  This peculiar velocity field $ {\bf v}({\bf r})$ can be related to the matter density contrast $\delta$ by

\begin{equation}
 {\bf v}({\bf r})=\frac{f_g H_{0}}{4 \pi} \int d^3 {\bf r}' \delta({\bf r'})  \frac{({\bf r}'-{\bf r})}{\left| {\bf r}'-{\bf r} \right|^3 }
 \label{peculair_velocity_matter}
\end{equation}
where $f_g$ is the growth rate of the density contrast, $\partial \ln \delta / \partial \ln a$, and $a$ is the scale factor \citep{0691019339}.  The density contrast is defined in terms of the ratio of the density at ${\bf r}$, $\rho$, to the average density $\bar \rho$, so that $\delta = \rho / \bar \rho$.

We can measure the peculiar velocity via the effect it has on the redshift of the galaxy.  The redshift $z$ of a galaxy is given by a contribution from the Hubble flow, $H_0r$, and the line of sight component of the peculiar velocity $S$:
\begin{equation}
cz=H_0r+S
\end{equation}
where, for a galaxy labelled $m$, at position ${\bf r}_m$, we have the line of sight peculiar velocity, $S_m$ given by
\begin{equation}
S_m=\hat{{\bf r}}_m \cdot {\bf v}({\bf r}_m)
\end{equation}
where $\hat{{\bf r}}_m$ is a unit vector in the direction of galaxy $m$.  To measure $S$ we thus need to combine the redshift of the galaxy with an \emph{independent} measure of the distance $r$.  In principle, we can calculate the peculiar velocity of any galaxy for which we have the luminosity distance, measured using distance indicators such as supernovae (SN), Tully-Fisher (TF) \citep{1977A&A....54..661T} or Fundamental-Plane (FP) \citep{1987ApJ...313...42D} measurements.  Individual uncertainties on luminosity distances are typically rather large (5\% for SN, and around 10 to 20\% for TF and FP), which propagates to a very large uncertainty in the peculiar velocity.  

\subsection{Maximum Likelihood Catalogue Method}

We now consider how to relate a catalogue of peculiar velocity measurements $S_m$ to large scale structure, following the method as presented in \cite{1995ApJ...455...26J}.  The method is based on a likelihood framework, where the likelihood $\mathcal{L}$ is given by
\begin{equation}
\mathcal{L}=\frac{1}{2\pi^{\nicefrac{N}{2}} \left| R_{mn} \right|^{ \nicefrac{1}{2}  } } \exp\left( -\frac{1}{2}S_mR^{-1}_{mn}S_n \right) 
\label{velocity_Likelihood}
\end{equation}
where $R_{mn}$ is the covariance matrix for the peculiar velocity measurements, for a catalogue of $N$ galaxies.  The covariance matrix is split into two components, a `velocity' and an `error' term: 
\begin{equation}
R_{mn}=R_{mn}^{(v)}+R_{mn}^{(e)} \;.
\label{CM_split}
\end{equation}
$R_{mn}^{(v)}$ models the coherent large scale structure, and $R_{mn}^{(e)}$ is a noise component to account for nonlinear velocity dispersion $\sigma_*$, and uncertainty in each peculiar velocity measurement $\sigma_m$.  If we assume that the measurement errors are uncorrelated, $R_{mn}^{(e)}$ simply contributes these uncertainties to the diagonal of the covariance matrix, and is given by
\begin{equation}
R_{mn}^{(e)}=(\sigma^2_m + \sigma^2_*)\delta_{mn}
\label{error_CM}
\end{equation}
where $\delta_{mn}$ is a Kronecker delta.  This formalism does not include correlations between measurement uncertainties.  In principle, there may be small correlations between measurement uncertainties due to, for example, galactic extinction, or the assumed mass-to-light ratio within catalogues of the same type of galaxy.  However, we assume that the cross-correlations in the data covariance matrix are dominated by the geometry of the survey, which we model with the  `velocity' component of the covariance matrix.  This term of the covariance matrix contains the power spectrum we are fitting for, and is given by
\begin{equation}
R^{(v)}_{mn}(k)=\int\frac{4\pi k^ 2 dk}{(2 \pi)^3}P_v(k)f_{mn}(k)
\label{velocity_CM}
\end{equation}
where $P_v(k)$ is the velocity power spectrum, which is related to the matter power spectrum by 
\begin{equation}
P_v(k)=\left( \frac{H_0 a}{k} \right) ^2 f_g^2P(k) \;.
\label{velocity_to_matter_Pk}
\end{equation}
Here $a$ is the scale factor.  The window function $f_{mn}(k)$ is given by
\begin{equation}
f_{mn}(k)=\hat{r}_{m,i}\hat{r}_{n,j}\int\frac{d^2\hat{k}}{4 \pi}\hat{k}_i \hat{k}_j e^{ ik {\bf \hat{k}} \cdot  (\bf{r}_m-\bf{r}_n)}
\label{f_matrix}
\end{equation}
and can be calculated analytically in terms of trigonometric functions.  A derivation of $f_{mn}$ is presented in \cite{2010arXiv1010.4276M}.  The general approach we take is to map out the likelihood in terms of parameters of the power spectrum.  We next consider an analogous method to analyse moments of the velocity field, before considering parametrisation of the power spectrum any further, since the parametrisation is common to both methods.

\subsection{Minimum Variance Moments Method}

\begin{figure}
\centering

    \includegraphics[width=9cm]{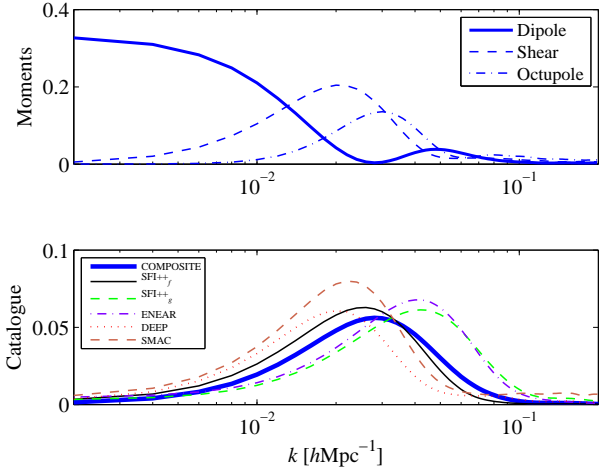}

\caption{Comparing sensitivity to different scales for the moments and catalogue methods.  For the moments panel, we plot the average of the diagonal of the covariance matrix for the dipole, shear and octupole terms, for COMPOSITE. Since the moments are designed to be orthogonal, the diagonal of the covariance matrix gives an indication of the scales at which the moments are most sensitive.  For the catalogue panel, the diagonal of the covariance matrix is $1/3$, so we plot the sum of the entire covariance matrix, normalised by the square of the number of galaxies in each catalogue, which illustrates the scales at which the window function is sensitive.  The sub-catalogues which span the greatest distance, DEEP and SMAC, are most sensitive to largest scales, while the shallowest catalogues, SFI++$_g$ and ENEAR, are most sensitive to smaller scales.}
    \label{WindowFunctionPlot}

\end{figure}

In this paper we directly compare results between the maximum likelihood catalogue method to an alternative method studying moments of the velocity field, from \cite{2011MNRAS.tmp..391M}.  We can consider the velocity field as a Taylor expansion, given by: 
\begin{equation}
v_{i}({\bf r})=U_{i}+U_{ij}r_{j}+U_{ijk}r_{j}r_{k}+\dots
\label{velocity_field_taylor_expansion}
\end{equation}
where $U_{i}$ is the dipole moment of the velocity field (often called the `bulk flow'), and provides most information about the largest scale fluctuations.  $U_{ij}$ is the shear of the velocity field, sensitive to intermediate scales.  
 $U_{ijk}$ is the octupole moment, and is sensitive to scales less than the size of the survey.  The relative sensitivity of each moment is shown in Figure \ref{WindowFunctionPlot}.  Since we can only measure the line of sight component of each peculiar velocity, each individual measurement must be weighted according to the component of the moment it is sensitive to.  \cite{2009MNRAS.392..743W} and \cite{2009arXiv0911.5516F} developed new `minimum variance' weights to estimate the velocity of the \emph{volume} traced by the galaxies in the survey.  These weights are designed to minimise the effects of small scale motions, to provide a better estimate of the large scale velocities.  We do not reproduce the derivations of the weights here; the method is presented in \cite{2009MNRAS.392..743W} for the dipole and extended in \cite{2009arXiv0911.5516F} for the shear and octupole.
  
With this method, the data consists of the three components of the dipole vector, six independent components of the shear, and ten independent components of the octupole.  We take a similar approach as with the catalogue method, splitting the covariance matrix into `velocity' and `error' terms.  The `velocity'  term is now given by
\begin{equation}
R^{(v)}_{pq}=\frac{f_g^2}{2 \pi^2}   \int dk P(k) \mathcal{W}^{2}_{pq}(k)
\label{velocity_matrix}
\end{equation}
where $p$ and $q$ index the 19 independent moments.  The window function $\mathcal{W}^{2}_{pq}(k)$ is sensitive to different scales for the dipole, shear and octupole, as plotted in Figure \ref{WindowFunctionPlot}.  We construct a likelihood in exactly the same manner as the catalogue method, in terms of parameters of a power spectrum, which we now consider.

\subsection{Power Spectrum Parametrisation}

\begin{figure}
\centering
\includegraphics[width=9cm]{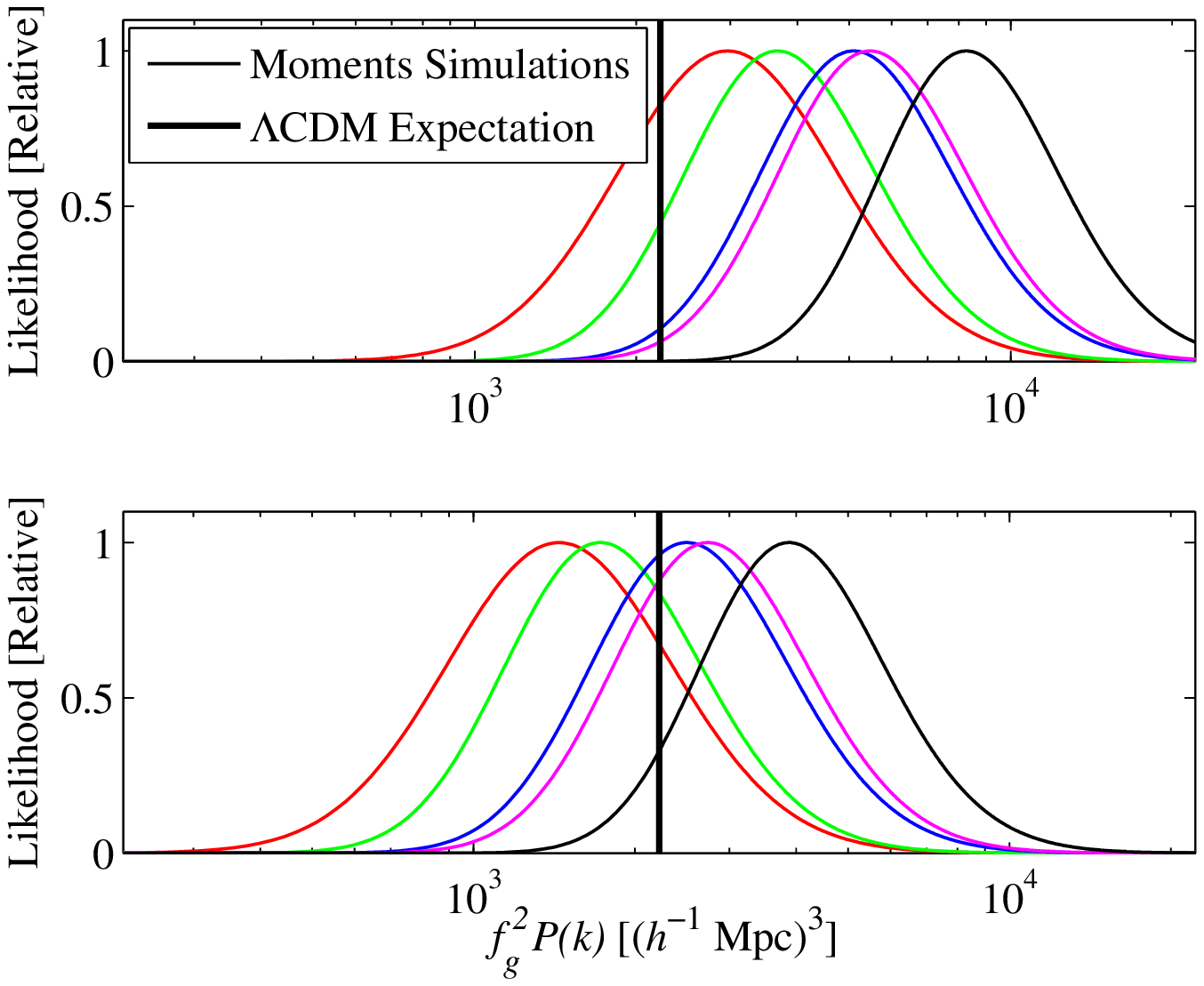}
\caption{The effect of band shape.  In the upper Figure, when using the constant amplitude of a single flat band-power to parametrise the power spectrum, we find a systematic excess compared to the average \lcdm power spectrum in the same $k$ range. However, when we use the amplitude of the \lcdm power spectrum, $A_{\alpha}$ as a parameter, we find no systematic deviation from the $A_{\alpha}=1$ expectation.  The results are for five different realisations of a \lcdm simulation, each consisting of dipole, shear and octupole moments.}
\label{band_shape_effect}

    \includegraphics[width=9cm]{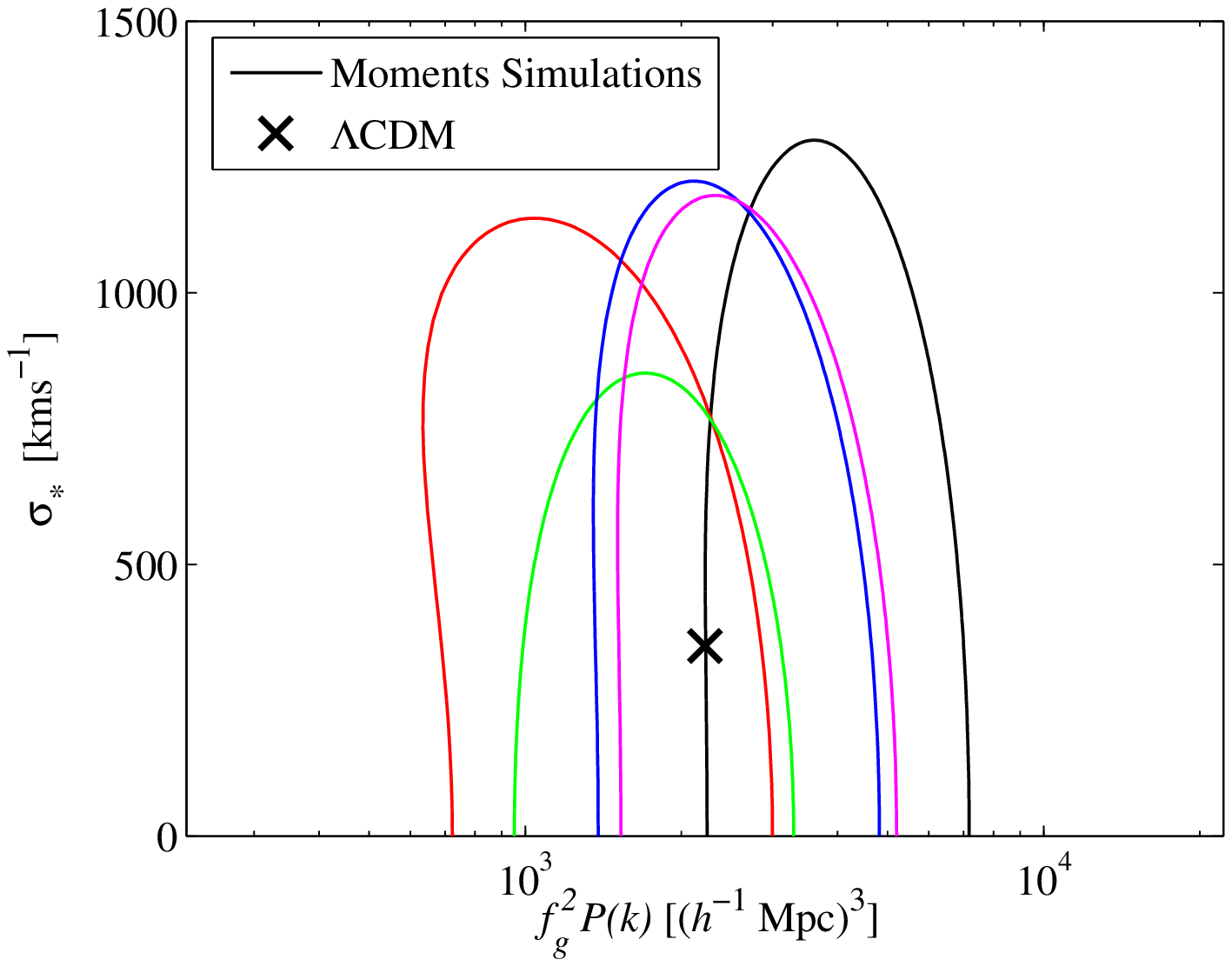}

\caption{One standard deviation contours for one power spectrum band and the velocity dispersion $\sigma_*$, for moments generated from a simulated velocity field.  The contours are fairly insensitive to the choice of $\sigma_*$.  Since compressing the velocity field into moments is insensitive to small scale effects, these contours are consistent with $\sigma_*=0$.  The 1d likelihood distributions are for a fixed  $\sigma_*$ of 200 kms$^{-1}$}
    \label{A_sigma_star_plot}
\end{figure}

Ultimately, we wish to constrain parameters of the underlying matter power spectrum.  There are many choices for ways to parametrise the power spectrum.  A popular choice in many earlier works was the power spectrum shape parameter $\Gamma$, the matter density $\Omega_m$ and the power spectrum normalization at 8 $h^{-1}$Mpc, $\sigma_8$ \citep{1995ApJ...455...26J,1995astro.ph.12132K}.  \cite{2001ApJ557102S} opted for a band-power parametrisation, which is the approach we take here, in terms of band-powers, $P_{\alpha}$, so the power spectrum is given by
\begin{eqnarray}
P(k)=\left\{\begin{array}{cl} P_{\alpha} &  k_{\alpha} < k < k_{\alpha+1} \\
0 & \mbox{otherwise.} \end{array}\right.
\end{eqnarray}
The velocity covariance matrix is then
\begin{equation}
R^{(v)}_{mn}(k) \approx \frac{(H)^2 }{2 \pi^2} f_g^2 \sum_{\alpha} P_{\alpha} \mathcal{K}_{\alpha}
\end{equation}
where $\mathcal{K}$ is given by
\begin{equation}
\mathcal{K}_{\alpha}=\int_{k_{\alpha}}^{k_{\alpha+1}} dk f_{mn}(k)
\end{equation}
This parametrisation is directly sensitive to the combination of $f_g^2 P_{\alpha}$.  To directly constrain $P(k)$, we must either assume a fiducial growth rate, or marginalise over other measurements of the growth rate.  For generality, we choose to treat the combination of $f_g^2P_{\alpha}$ as a parameter.

In \cite{2011MNRAS.tmp..391M}, we used flat band-powers.  This parametrisation was sufficient to demonstrate the large effect of including just the dipole, or higher moments of the velocity field.  However, in the widest, single band parametrisation, the flat bands can introduce an apparent artificial shift towards an excess of power, shown in the upper panel of Figure \ref{band_shape_effect}.  We find that when we factor in a fiducial \lcdm power spectrum into the window function, $P_{\textrm{fid}}(k)$, and allow a constant amplitude of this power spectrum, $A_{\alpha}$, to vary, we obtain an amplitude which is consistent with $\Lambda$CDM.  That is, for the same average amplitude of $P(k)$ the most likely value can depend on the slope of the band within the band range.  We find in practice that this shift is most significant for the single band parametrisation, and when we allow several bands to vary across the range the slope of each band is less significant.  Nevertheless, for consistency and completeness we use \lcdm shaped bands for all parametrisations, varying the average amplitude within each band.  The kernel in this case is
\begin{equation}
\mathcal{K}_{\alpha}=\int_{k_{\alpha}}^{k_{\alpha+1}} dk f_{mn}(k)P_{\textrm{fid}}(k)
\end{equation}
so our velocity covariance matrix is now 
\begin{equation}
R^{(v)}_{mn}\approx \frac{(H)^2 }{2 \pi^2} f_g^2 \sum_{\alpha} A_{\alpha} \mathcal{K}_{\alpha}
\label{velocity_CM}
\end{equation}
where our parameter is now $A_{\alpha}$, the amplitude of the fiducial power spectrum in each band.  We thus expect $A_{\alpha}=1$ in all bands for a \lcdm velocity field.  To more easily compare $A_{\alpha}$ to expectations, we multiply the results by the mean of the power spectrum in each band, $\left<P_{\alpha}\right>$, given by
 \begin{equation}
\left<P_{\alpha}\right>= \frac{\int_{k_{\alpha}}^{k_{\alpha+1}} dk P_{\textrm{fid}}(k)}{k_{\alpha+1} - k_{\alpha}} \;.
\label{P_alpha_def}
\end{equation}
Although this parametrisation specifies the shape of the power spectrum within each band, we see in Section \ref{Results} in simulated catalogues that the method is fairly robust to the choice of fiducial power spectrum.  We find that the catalogue method is sensitive to the velocity dispersion parameter, and thus treat it as a free parameter in the manner of  \cite{2010arXiv1010.4276M}.  To directly compare between the catalogue and moments method, we repeat much of the analysis of \cite{2011MNRAS.tmp..391M} with $\sigma_*$ as a free parameter, although we find with results from velocity moments that marginalising over the velocity dispersion parameter has only a very small effect.  The same set of moments from Figure \ref{band_shape_effect} are analysed in Figure \ref{A_sigma_star_plot} allowing both the velocity dispersion and amplitude to vary.  In this paper we now consider the combination of $f_{g}^{2} P_{\alpha}$ as a parameter for both methods, as opposed to marginalising over the growth rate as in  \cite{2011MNRAS.tmp..391M}.  We assume a fiducial value for the velocity dispersion of 350 km s$^{-1}$.

We use a Metropolis-Hastings Markov-Chain Monte-Carlo method to map out the likelihood expression in terms of the band-powers and velocity dispersion.  The error bars plotted on the power spectrum plots are calculated at the one standard deviation confidence level from the marginalised likelihood distribution for each band-power.  Where the likelihood peaks at zero this is plotted without a marker, and the error bar represents the extent of the upper bound.

\section{Peculiar Velocity Data}
\label{PeculiarVelocityData}

\begin{figure*}
\begin{center}
\epsfxsize 19cm
\epsfysize 7cm
\epsfbox{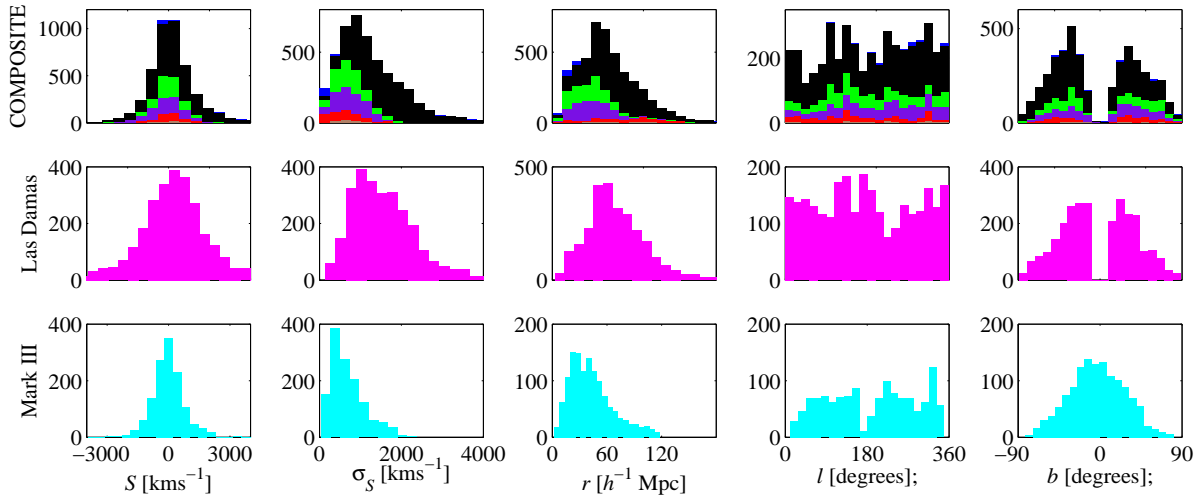}
\caption{Histograms of the real surveys, and the Las Damas and Mark III simulated surveys, illustrating the distribution of line of sight peculiar velocities, $S$, uncertainty in $S$, $\sigma_S$, distance $r$, and coordinates $l$ and $b$.  The COMPOSITE survey is plotted as a stacked histogram and is comprised of the SFI++$_{f}$ (black), SFI++$_{g}$ (green), ENEAR (purple), DEEP (red), and SMAC (brown) surveys.}
\label{Survey_histograms}
\end{center}
\end{figure*}

We primarily study the COMPOSITE peculiar velocity catalogue, which consists of 4537 individual peculiar velocity measurements, and a characteristic depth of 33 $h^{-1}$Mpc, compiled by \cite{2009arXiv0911.5516F}.  The COMPOSITE catalogue is composed of several sub-catalogues, which we also consider individually.  The largest sub-catalogue is the SFI++ (Spiral Field I-band) sample, which consists of 3456 TF measurements.  We analyse the field galaxies and groups in the SFI++ sample separately, as SFI++$_{f}$ (2720 galaxies) and SFI++$_{g}$ (736 measurements) \citep{2006ApJ...653..861M,2009ApJS..182..474S,2009yCat..21720599S}.  We also study the combined DEEP catalogue, compiled by \cite{2009MNRAS.392..743W}, which consists of 294 of the deepest peculiar velocity measurements and the ENEAR (Early-type NEARby galaxies) catalogue \citep{2000ApJ...537L..81D}.  The smallest catalogue we consider individually is the SMAC catalogue \citep{2004MNRAS.352...61H}, which is a sub-set of the DEEP catalogue.

We test our procedures on peculiar velocity catalogues generated from \lcdm simulations.  We analyse the set of 20 simulated Mark III catalogues\footnote{downloaded from \url{http://www.mpa-garching.mpg.de/NumCos/CR/Download/index.html}} drawn from realisations of the Virgo simulation \citep{1996ApJ...458..419K}, consisting of 1300 entries each (some of the catalogues consisted of slightly more than 1300 entries - these were trimmed to 1300 for consistency).  We also study six simulated catalogues from the Las Damas simulation (McBride et al., in prep.), designed to resemble the SFI++$_{f}$ catalogue with 2720 entries.  Histograms of the real and simulated surveys are shown in Figure \ref{Survey_histograms}.

We study the simulated data sets with both the catalogue and moments method.  We analyse the real data with the catalogue method for the combined COMPOSITE set, and the five sub-catalogues.  We find that the sub-catalogues are too small to reliably study the moments alone, so we present results here with the moments method only for the COMPOSITE catalogue.

\section{Results}
\label{Results}

\subsection{Simulated Data}
  
\begin{figure}
\centering

    \includegraphics[width=9cm]{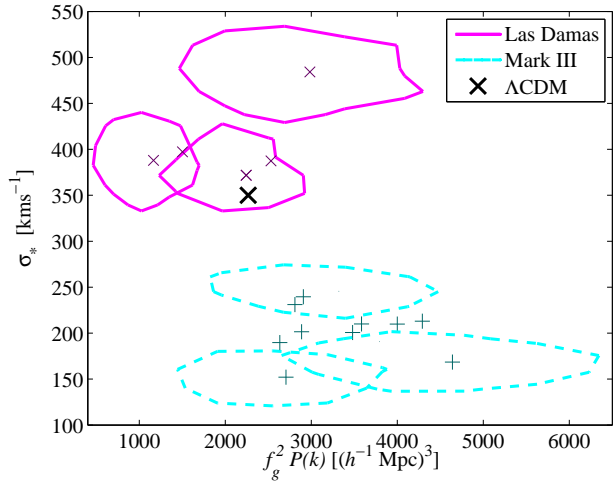}

\caption{Constraints on the velocity dispersion parameter $\sigma_*$ and one band-power, in the range $k=0.002$ to $0.196$ $h^{-1}$Mpc, for \emph{simulated} peculiar velocity surveys from the Las Damas and Mark III simulations. The markers represent the peak of the likelihood for each catalogue.  The contours are 1 standard deviation uncertainty about the peak likelihood, which we have only plotted for three catalogues for each simulation set, to prevent the plot from becoming overcrowded.  Similarly, we only plot results for 10 of the Mark III catalogues.  These uncertainty ranges are typical for the points plotted here. }
\label{ContourPlot_1Band_LasDams_MarkIII} 
\end{figure}

We begin by testing our method on peculiar velocity catalogues generated from \lcdm simulations.  We start the analysis with the simplest parametrisation: one band-power.  We choose a window function range of $k=0.002$ to $0.196$ $h^{-1}$Mpc, to match the $k$ range used in \cite{2011MNRAS.tmp..391M}.  We also include the velocity dispersion as a free parameter.  Results are shown in Figure \ref{ContourPlot_1Band_LasDams_MarkIII}.

\begin{figure}
\centering

    \includegraphics[width=9cm]{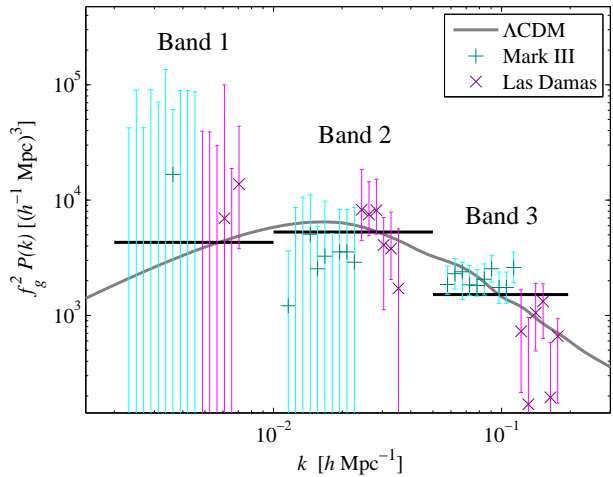}

\caption{Results from the Mark III and Las Damas mocks in three band-powers.  The velocity dispersion was varied as a free parameter, and marginalized over in these results.  The grey curve is the \lcdm power spectrum.  The horizontal black lines span the $k$ ranges of the band-powers, and indicate the average value of the power spectrum across this range.  The results should be compared directly to these.  The markers are the marginalized results for each band-power, for each of the Mark III and Las Damas surveys.  The $k$ location within each band is arbitrary; each point should be fully considered equally at the centre of the band, spanning the full width.}
\label{PkPlot_ThreeBand_Sim_MIII_LasDamas} 
\end{figure}

We find that for this band-power parametrisation, the velocity dispersion is uncorrelated with the power spectrum amplitude.  The velocity dispersion is slightly higher than the \lcdm expectation for the Las Damas mocks, although it is lower in the Mark III mocks.  In both sets of mocks, we note a small systematic shift from the \lcdm expectation.  As this shift is in opposite directions for the Mark III and Las Damas mocks, it seems reasonable to conclude that this may be due to the many particularities of generating a mock catalogue from an underlying power spectrum, as opposed to a systematic shift introduced by our method.  In other words, the Mark III and Las Damas mocks taken together surround the \lcdm expectation well.  Following \cite{2011MNRAS.tmp..391M}, we next consider a parametrisation with three band-powers, spanning the same $k$ range with bands evenly spaced in $\log k$.  The results are shown in Figure \ref{PkPlot_ThreeBand_Sim_MIII_LasDamas}.

\begin{figure}
\centering

    \includegraphics[width=9cm]{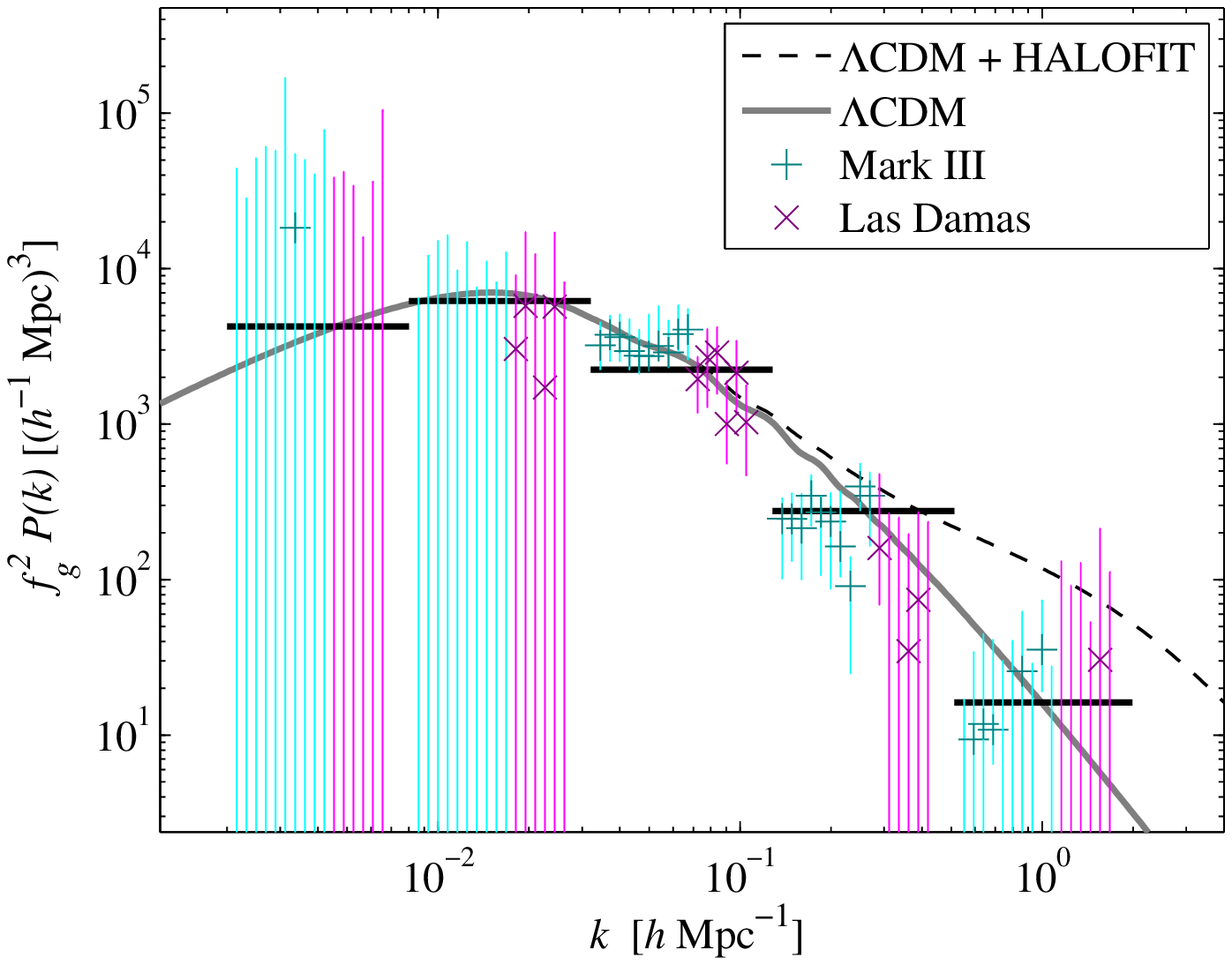}

\caption{As Figure \ref{PkPlot_ThreeBand_Sim_MIII_LasDamas}, but with two additional bands added at smaller scales.  The smallest scale band is at scales for which galaxy power spectra must be corrected for (e.g., with Halofit).  With the peculiar velocity method, we are sensitive to the total matter distribution, and can thus directly probe the \emph{linear} power spectrum, without accounting for halo corrections.}
\label{PkPlot_FiveBand_sim_NoHALOFIT} 

\centering

    \includegraphics[width=9cm]{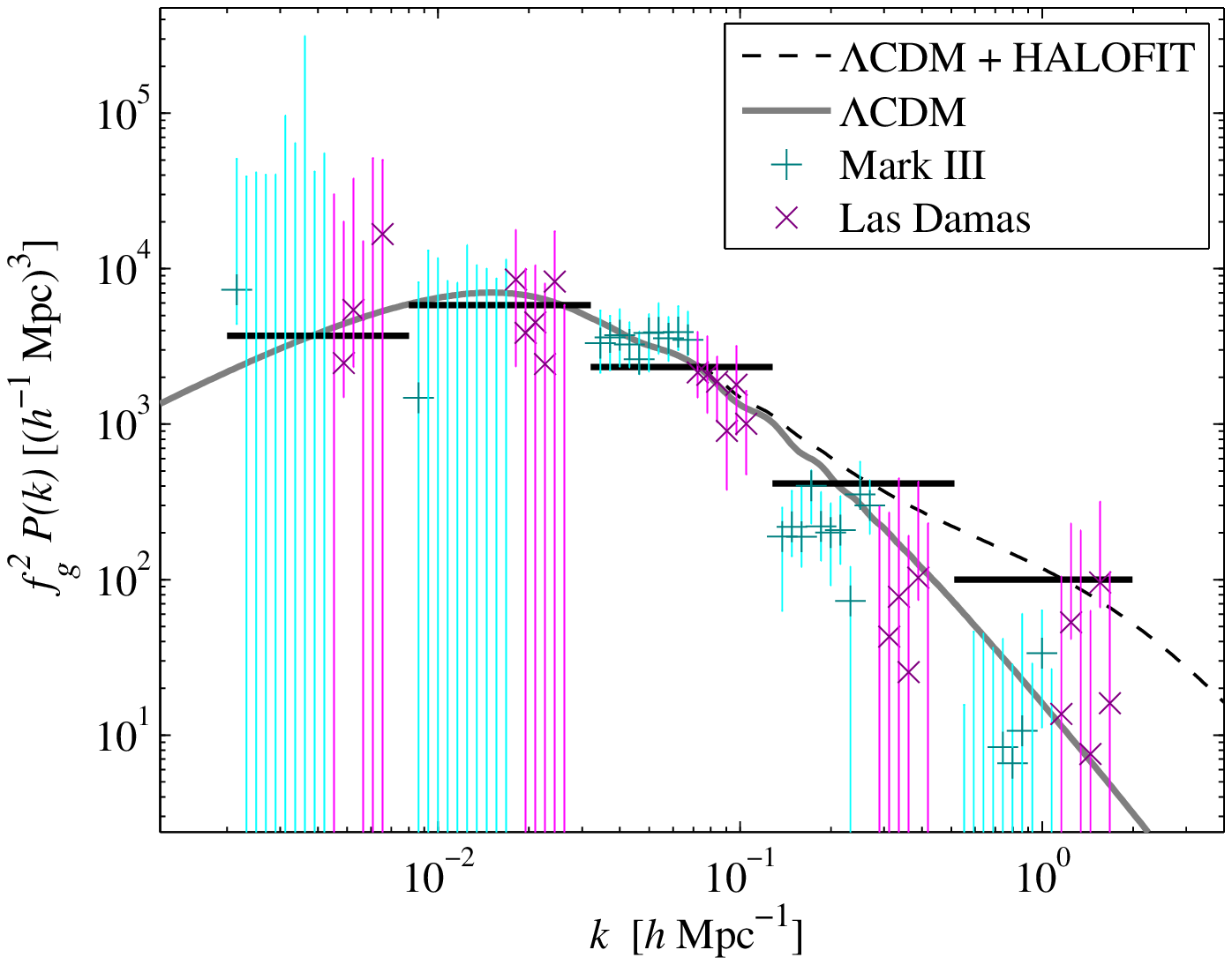}

\caption{As Figure \ref{PkPlot_FiveBand_sim_NoHALOFIT}, but with a fiducial power spectrum corrected with Halofit.  The black bars now represent the mean value of the Halofit fiducial power spectrum which was factored into the window function.  The data still favor the linear power spectrum, indicating that the method is robust to the choice of fiducial power spectrum.}
\label{PkPlot_FiveBand_sim_HALOFIT_fiducialPk} 
\end{figure}

We will refer to the bands from the largest scales to the smallest scales as bands 1 to 3.  We find that the uncertainty in band 1 is larger in the simulated data with the catalogue method than the moments method (not plotted). The uncertainty in band 1 is particularly important for understanding the high dipole moment, and will be discussed further in Section \ref{DiscussionAndConclusions}.  Both sets of catalogues are much more sensitive in band 3 than with the moments method; we thus decide to extend that band-power parametrisation by adding two more bands (4 and 5) at smaller scales.  Results are shown in Figure \ref{PkPlot_FiveBand_sim_NoHALOFIT}.

We find that band 5 is anti-correlated with the velocity dispersion parameter.  This is not surprising, given the $\sim5$ $h^{-1}$Mpc scales spanned by this band.  We are encouraged to see a difference at over $1\sigma$ from the nonlinear Halofit \citep{2003MNRAS.341.1311S} corrections to the galaxy power spectra, illustrating the ability of peculiar velocity measurements to directly probe the underlying \emph{linear} matter distribution.  As a test, we repeat the results in Figure \ref{PkPlot_FiveBand_sim_NoHALOFIT} assuming a fiducial Halofit corrected power spectrum to factor into the window function.  The results are shown in Figure \ref{PkPlot_FiveBand_sim_HALOFIT_fiducialPk}.  Encouragingly, we see in the smallest two bands that the data still favour the linear power spectrum.

\subsection{Real Data}
\begin{figure}
\centering
\includegraphics[width=9cm]{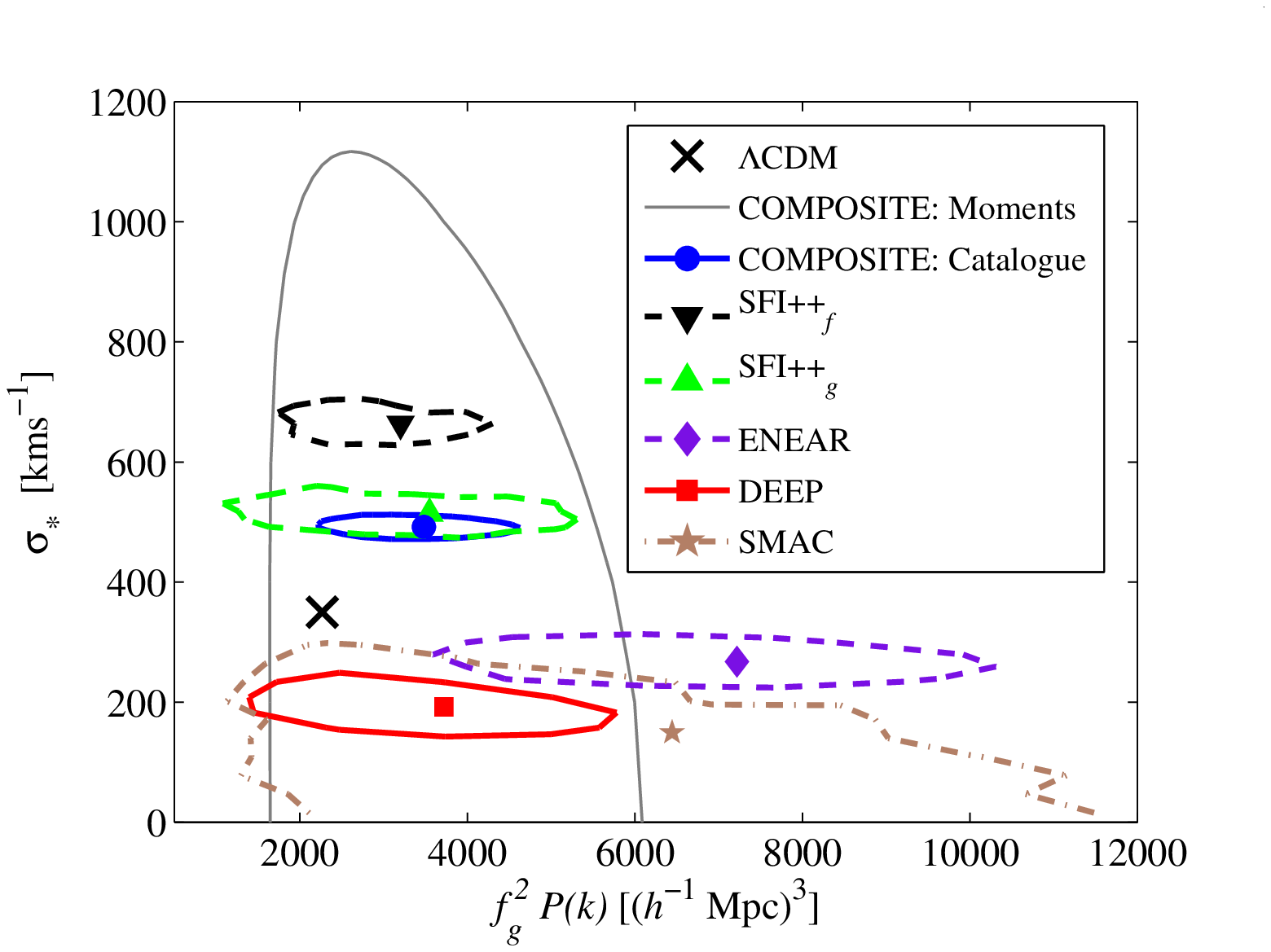}
\caption{1 standard deviation contours from the real data, parametrised in terms of the velocity dispersion $\sigma_*$ and one band-power, in the range $k=0.002$ to $0.196$ $h^{-1}$Mpc.  There is a slight excess of power in all the catalogues, although within the $1\sigma$ level for most.}
\label{ContourPlot_1Band_Real} 
\end{figure}

We now consider results from real peculiar velocity surveys.  We start with the one band-power and velocity dispersion parametrisation, shown in Figure \ref{ContourPlot_1Band_Real}.

\begin{figure}
\centering
\includegraphics[width=9cm]{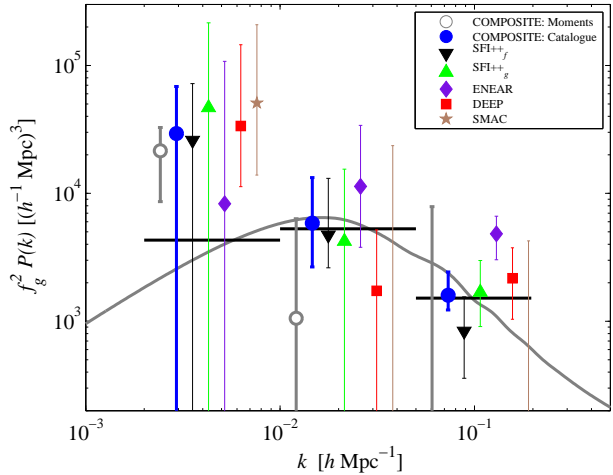}
\caption{As Figure \ref{PkPlot_ThreeBand_Sim_MIII_LasDamas}, for the real peculiar velocity catalogues.  As before, the $k$ position of each point within the bin is arbitrary, and each point should be fully considered at the centre of each band. The points have been ordered in each band from left to right by the number of galaxies in each catalogue.  We also plot results from the moments analysis of the COMPOSITE catalogue.}
\label{PkPlot_ThreeBand_Real} 
\end{figure}

Using the \lcdm shape band, we find a slightly lower average amplitude than the flat band result from \cite{2011MNRAS.tmp..391M}.  Although we note a small excess, the \lcdm value is now well within the 1$\sigma$ uncertainty range of the moments result.  We find that the results for the velocity dispersion agree well with similar results from  \cite{2010arXiv1010.4276M}.  We find a smaller uncertainty in the band-power with the catalogue method than the moments method, and a much improved constraint on the velocity dispersion parameter.  We next consider a three band parametrisation, shown in Figure \ref{PkPlot_ThreeBand_Real}.

\begin{figure}
\centering
\includegraphics[width=9cm]{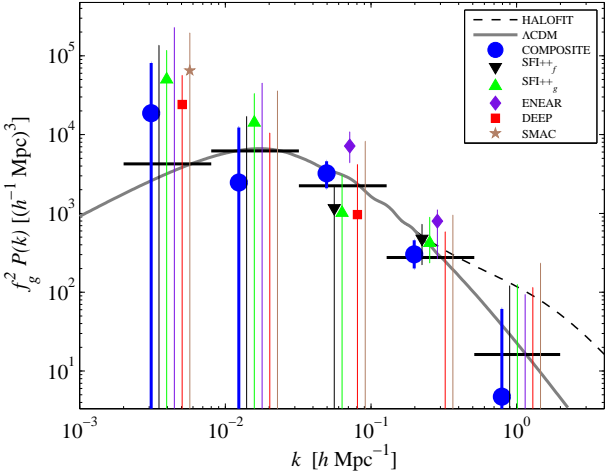}
\caption{As Figure \ref{PkPlot_FiveBand_sim_NoHALOFIT}, now for real data.  In band 1 we see a slight excess of power, although the uncertainty includes the \lcdm model at the 1$\sigma$ level.  Band 5 favours the linear clustering power spectrum over the nonlinear model.}
\label{PkPlot_FiveBand_Real_Updated} 
\end{figure}

With the moments method, band 3 was merely an upper limit - it is well constrained with the catalogue method.  The low shear of the velocity field caused band 2 to be lower than the \lcdm expectation.  When we now analyse the full catalogue, the band agrees extremely well with the \lcdm expectation.  We still observe an excess of power in band 1, although the lower bound is not constrained here.  These results are discussed further in Section \ref{DiscussionAndConclusions}.  We next extend the parametrisation with two smaller bands, as before with the simulated catalogues.  The results are plotted in Figure \ref{PkPlot_FiveBand_Real_Updated}, and presented for COMPOSITE in Table \ref{CompResultsTable}.  We find good agreement with \lcdm in bands 3 and 4.  We find that bands 1 and 2 are anti-correlated.

\begin{table}
\begin{center}
\begin{tabular}{c | c c}
 & \multicolumn{2}{|c|}{$f_g^2P(k)$  [$(h^{-1}$Mpc$)^3$]} \\          
Band & \lcdm & COMPOSITE  \\
\hline
1 &$4.2\times10^3 $&$1.9 ^{+6.2}_{-1.9} \times 10^4$ \\
2 &$6.2\times10^3 $&$2.5 ^{+9.9}_{-2.5} \times 10^3$ \\
3 &$2.2\times10^3 $&$3.2^{+1.4}_{-1.1}  \times 10^3$\\
4 &$2.8\times10^2$ &$3.0^{+1.5}_{-1.0}  \times 10^2$\\
5 &$1.6\times10^1 $&$0.5^{+5.7}_{-0.5}  \times10^1$\\
\end{tabular}
\end{center}
\caption{Results for the COMPOSITE catalogue five-band parametrisation, as plotted in Figure \ref{PkPlot_FiveBand_Real_Updated}.  Although the high dipole velocity leads to an excess of power by over a factor of 4 in band 1, the uncertainty includes the \lcdm model within the 1$\sigma$ uncertainty level.}
\label{CompResultsTable}
\end{table}

\subsubsection{Comparison to \protect\cite{2001ApJ557102S}}
\begin{figure}
\centering
 \includegraphics[width=9cm]{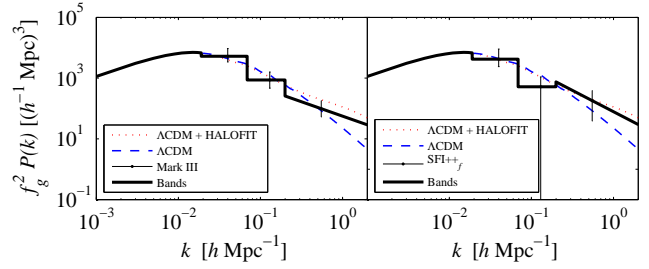}
\caption{Analyzing a simulated Mark III catalogues with the parametrisation of \protect\cite{2001ApJ557102S}.  The results we find here agree well with the results in Figure 7 of of \protect\cite{2001ApJ557102S}.}
\label{PlotSilberman} 
\end{figure}

We also test our methodology by reproducing the results of \cite{2001ApJ557102S}, to which our method is similar.  We change our bands as follows, to match the bands used in their work:  For $k \leq 0.02 $ $ h$Mpc$^{-1}$ we use a \lcdm band, which we do not vary.    In the range $0.02 < k \leq 0.07$ and $0.07 < k \leq 0.2 $ $h$Mpc$^{-1}$ we use two independent flat bands of constant amplitude.  For $k > 0.02 $ $h$Mpc$^{-1}$ we use a power law band, $A_{\alpha}k^n$  The spectral slope $n$ is set to -0.95 for the simulated Mark III catalogues, and to -1.4 for the real SFI++$_f$ catalogue, and the free parameter $A_{\alpha}$ is varied. The results are shown in Figure \ref{PlotSilberman}.  We reproduce the slight decrement at $k\sim0.1$ $ h$Mpc$^{-1}$ as noted by \cite{2001ApJ557102S}, which we also observe in the SFI++$_f$ catalogue in band 3 of our 5 band parametrisation in Figure \ref{PkPlot_FiveBand_Real_Updated}.

\subsection{Fitting a Velocity Dipole}

\begin{figure}
\centering
\includegraphics[width=9cm]{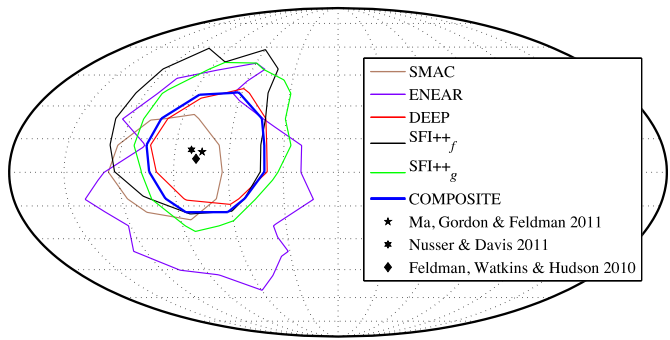}
\caption{The direction of the best-fitting velocity dipole in the catalogues, (in a Mollweide projection of the sky in galactic coordinates).  The contours represent 1 standard deviation levels on the dipole direction angles $U_l$ and $U_b$ found in the peculiar velocity catalogues. The direction agrees with other results from \protect\cite{2009arXiv0911.5516F}, \protect\cite{2010arXiv1010.4276M}  and \protect\cite{0004-637X-736-2-93}. } 
\label{DipoleMap}
\end{figure}

In the three band model, we find that the uncertainty in band 1 is larger with the catalogue method than with the moments method.  We see these results in the simulated catalogues, but it is particularly important to understand the result in the COMPOSITE catalogue since it is closely related to the high velocity dipole.  As can be seen in Figure \ref{WindowFunctionPlot}, the catalogue method window function is most similar to the window function of the shear moment, and (relatively) not as sensitive to the scales probed by band 1.  To try and understand these results, we use our maximum likelihood method to estimate the dipole, to compare to the minimum variance dipole of \cite{2009arXiv0911.5516F}.  Instead of varying the power spectrum, we fix the power spectrum at the fiducial \lcdm value and minimise parameters of the dipole, as analysed by \cite{2010arXiv1010.4276M}.  We find that it is not possible to simultaneously constrain the three band-power spectrum and a velocity dipole.  Following \cite{2010arXiv1010.4276M}, we model the dipole ${\bf U}$ by subtracting the line of sight component of the dipole to each galaxy in the catalogue from it's line of sight peculiar velocity, $S_n$, to obtain a `tilted' velocity, $p_n$ 
\begin{equation}
p_n = S_n - \hat{r}_n \cdot U \;.
\end{equation}
We parametrise the dipole bulk flow as ${\bf U}=\{ U_r, U_l, U_b  \}$, where $U_r$ is the magnitude of the bulk flow (in \kms), and $U_l$ and $U_b$ are the direction of the flow in degrees (galactic coordinates).  We also include the velocity dispersion $\sigma_*$ as a free parameter, which we marginalise over to find the dipole bulk flow.  We find a dipole of magnitude $U_r=380^{+99}_{-132} $ km s$^{-1}$ in the direction $U_l=295^{+18}_{-18} $ and $U_b=14\pm18 $ degrees.  The direction of the dipole is shown in Figure \ref{DipoleMap}.  This is slightly slower and with a larger uncertainty than the dipole moment found by \cite{2009arXiv0911.5516F} of $U_r=416\pm78 $ km s$^{-1}$ in the direction $U_l=282 \pm 11 $ and $U_b=6 \pm 6 $ degrees, and is related to the larger uncertainty we find in band 1 with the catalogue method than the moments method.  A comparison of measurements of the dipole is shown in Table \ref{DipoleMeasurements}.

Results from many different surveys and analysis methods appear to agree on the \emph{direction} of the velocity dipole, at around $l=280$ and $b=10$ degrees, within an uncertainty radius of about 15 degrees.  However, there is less consensus as to the magnitude of the dipole.  The magnitude of the dipole (and the corresponding uncertainty) can depend on the depth of the survey, the treatment of outliers, and the weighting of galaxies.  Even so, we can very conservatively state that the magnitude of the velocity dipole out to 100 \rUnit is a factor of several times higher than the \lcdm expectation.  

\begin{table*}
\begin{center}
{\small
\hfill{}
\begin{tabular}{r c c c}
 & $U_r$ [\kms] & $U_l$ [degrees]& $U_b$ [degrees] \\
\hline
\protect\cite{2009MNRAS.392..743W} & 407 $ \pm $ 81 & 287 $ \pm $ 9 & 8 $ \pm $ 6 \\
\protect\cite{2009arXiv0911.5516F} & 416 $ \pm $ 78 & 282 $ \pm $ 11 & 6 $ \pm $ 6 \\
\protect\cite{2010arXiv1010.4276M} & 340 $ \pm $ 130 & 285.1$^{+23.9}_{-19.5}$ & 9.1$^{+18.5}_{-17.8}$ \\
\protect\cite{0004-637X-736-2-93} (SFI++, 40 \rUnit) & 333 $ \pm $ 38 & 276 $ \pm $ 3 & 14 $ \pm $ 3 \\
(SFI++, 100 \rUnit) & 257 $ \pm $ 44 & 279 $ \pm $ 6 & 10 $ \pm $ 6 \\
This work & $380^{+99}_{-132}$  & 295$ \pm $ 18  & 14 $ \pm $18  \\
\end{tabular}}
\hfill{}
\caption{Comparison of dipole measurements within 100 $h^{-1}$Mpc.  This work, \protect\cite{2009MNRAS.392..743W}, \protect\cite{2009arXiv0911.5516F} and \protect\cite{2010arXiv1010.4276M} all study the COMPOSITE catalogue, whereas \protect\cite{0004-637X-736-2-93} study the SFI++ catalogue, which comprises the majority of COMPOSITE.  The direction of the flow from different works agree well, although there is considerable variation in the magnitude of the flow.  The magnitude of the flow can depend strongly on the depth of the survey, how galaxies at different depths are weighted, and the sensitivity of the method the fiducial power spectrum.}
\label{DipoleMeasurements}
\end{center}
\end{table*}

\section{Discussion \& Conclusions}
\label{DiscussionAndConclusions}

In terms of understanding the high dipole moment, the main difference between the two methods we have considered here is that the catalogue method finds an excess of power in band 1 with an uncertainty which includes the \lcdm model at the 1$\sigma$ level, while the moments method finds an excess of power with an uncertainty which excludes the \lcdm model at over the 1$\sigma$ level.  This is the issue we now consider.  The power spectrum at these scales is closely linked to the magnitude of the dipole moment, where we similarly find a larger uncertainty with the catalogue method than the minimum variance moments result.  The key to the different results are the minimum variance weights, which are designed to be sensitive to the largest scales, at the expense of small scale information.  Indeed, of the sub-catalogues we analyse with the catalogue method, band 1 is best constrained by the DEEP and SMAC samples, which are the deepest of the sub-catalogues considered here.  While both DEEP and SMAC are included within COMPOSITE, we find that the additional, shallower galaxies are effectively acting as noise as far as constraints on band 1 are concerned.  Essentially, the minimum variance weights achieve this effect to a maximal extent: preferentially weighting the deeper galaxies with the cleanest measurement of the velocity moments, at the expense of small scale information. 

On smaller scales, we also find a difference between the two methods in band 2.  We find that the catalogue method agrees extremely well with the \lcdm model, while the moments method underestimates the power.  This is due to the low shear of the velocity field, which provides most sensitivity at the scales of band 2.  When we analyse the full velocity catalogue, the small scale motions not modelled by the moments appear to combine to provide a constraint which again agrees extremely well with $\Lambda$CDM.  In much the same way as the one band parametrisation in \cite{2011MNRAS.tmp..391M} illustrated the difference of including only the dipole, or additionally the shear and octupole, the result in band 2 highlights the effect of considering only moments of the velocity field, or the full information available to us.

In the parametrisation we have chosen, we have imposed that $P(k)=0$ outside the range of the band-powers, which could, in principle, affect the results in the bands at the edge of the parametrisation.  However, for the five band parametrisation, the range of the bands is sufficiently wide that the entire sensitivity range of the catalogues is covered, and the bands are insensitive to power beyond these scales.  We can also directly assess this effect by comparing the three and five band models we have presented here.  In the three band model, we have that $P(k)=0$ at the scales where $P(k)\not=0$ in the five band model (when we include bands 4 and 5).  However, we find that the results in band 3 are similar for both models (as can be seen in figures \ref{PkPlot_ThreeBand_Sim_MIII_LasDamas} \& \ref{PkPlot_FiveBand_sim_NoHALOFIT} for the simulated catalogues, and Figures \ref{PkPlot_ThreeBand_Real} \& \ref{PkPlot_FiveBand_Real_Updated} for the real catalogues), whether the neighbouring band is zero (the three band model) or nonzero (the five band model).  

We can also consider the issue of whether our method is robust to systematic errors in the peculiar velocity data.  For example, the Tully-Fisher and Fundamental Plane measurements we consider both depend on the assumed mass-to-light ratio of the observed galaxies.   To quantify the effect of systematic errors in the distance indicators, we have re-analysed a simulated Mark III catalogue and the genuine SGI++$_g$ catalogue, where we have artificially introduced a systematic offset in the distance indicator and then repeated the power spectrum analysis.  This mimics a systematic error which may be introduced by incorrectly determining the intrinsic luminosity of the galaxies.  We have analysed each catalogue with eight different artificial offsets in the distance indicators in the range from an underestimate of the distance by a factor of 0.6 (corresponding to an overestimate of the peculiar velocity) to an overestimate of the distance uncertainty by a factor of 1.4 (corresponding to an underestimate of the peculiar velocity.  As expected, we find a consistent overestimate of the band-powers when the peculiar velocity is overestimated, and a consistent underestimate in the band-powers when the peculiar velocity is underestimated.  We find that the ultimate systematic error in the bands scales linearly as approximately twice the error of the artificial distance offset.  

However, a robust indication that such systematic errors in the distance measurements are not problematic is that we observe similar results in each of the sub-catalogues of COMPOSITE that we study.  The DEEP and SFI++ catalogues consist of different distance indicators, and we observe similar results both for the power spectrum fitting and dipole fitting; there is no individual catalogue which is anomalous.

Perhaps more problematic may be systematic errors which can introduce a dipole variation in the observed luminosities, such as galactic extinction.  Such a systematic effect would affect all the catalogues in the same manner, and could also mimic a bulk flow.  From analysing moments of the velocity field, we know that a velocity dipole has most effect on the large scales of bands 1 and 2; the results we find in bands 3, 4 and 5 should be fairly robust to any systematic errors which introduce a dipole variation.  As for bands 1 and 2, although some of the velocity dipole could, in principle, be due to galactic extinction, it would then be difficult to reconcile with the low shear of the velocity field, and even harder to mimic with systematic effects.  In other words, the high dipole and low shear, when considered together, are indicative of the large scale over densities we see in band 1, and are difficult to fully mimic with systematic effects.

To conclude, we find that inferring the underlying power spectrum from peculiar velocity catalogues continues the general trend of \cite{2011MNRAS.tmp..391M}: that including more detail in the velocity field improves the agreement with the \lcdm model.  Specifically, we observe good agreement with \lcdm on scales of $k > 0.01$ $ h$Mpc$^{-1}$, although the agreement with \lcdm in band 1 from the catalogue method is only due to the larger uncertainty than the moments method.  While the high dipole moment alone may appear anomalous, when we consider the full peculiar velocity measurements, we find a power spectrum which agrees well with the \lcdm model.

\section{Acknowledgments}
PGF acknowledges support from STFC, BIPAC and the Oxford Martin School. HAF has been supported in part by an NSF grant AST-0807326 and by the National Science Foundation through TeraGrid resources provided by the NCSA.  MJH has been supported by NSERC and acknowledges the hospitality of the Institut d'Astrophysique de Paris, and the financial support of the IAP/UPMC visiting programme and the French ANR (OTARIE). 

\vspace{-0.3cm}

\bibliographystyle{mn2e}
\bibliography{peculiar_velocity}

\end{document}